\documentclass[
    twocolumn,			
    reprint, 			
    superscriptaddress, 
    amsmath, 			
    amssymb, 			
    prl,				
    aps,				
    floatfix			
    ]{revtex4-2}

\usepackage{graphicx}	
\usepackage{braket}		
\usepackage{amsmath}    
\usepackage{dcolumn}	
\usepackage{bm}			
\usepackage{subdepth}	
\usepackage{gensymb}	
\usepackage{color,soul}

\begin{document}

\title{Single photon frequency conversion for frequency multiplexed quantum networks in the telecom band}
\author{Paul Fisher}
\email{paul.fisher@griffithuni.edu.au}
\affiliation{Centre for Quantum Computation and Communication Technology (Australian Research Council), Centre for Quantum Dynamics, Griffith University, Brisbane, QLD 4111, Australia}
\author{Robert Cernansky}
\affiliation{Centre for Quantum Computation and Communication Technology (Australian Research Council), Centre for Quantum Dynamics, Griffith University, Brisbane, QLD 4111, Australia}
\author{Ben Haylock}
\affiliation{Centre for Quantum Computation and Communication Technology (Australian Research Council), Centre for Quantum Dynamics, Griffith University, Brisbane, QLD 4111, Australia}
\author{Mirko Lobino}
\email{m.lobino@griffith.edu.au}
\affiliation{Centre for Quantum Computation and Communication Technology (Australian Research Council), Centre for Quantum Dynamics, Griffith University, Brisbane, QLD 4111, Australia}
\affiliation{Queensland Micro- and Nanotechnology Centre, Griffith University, Brisbane, QLD 4111, Australia}

\begin{abstract}
    High-speed long-range quantum communication requires combining frequency multiplexed photonic channels with quantum memories. We experimentally demonstrate an integrated quantum frequency conversion protocol that can convert between wavelength division multiplexing channels in the telecom range with an efficiency of $55\pm 8\%$ and a noise subtracted HOM dip visibility of 84.5\%. This protocol is based on a cascaded second order nonlinear interaction and can be used to interface a broad spectrum of frequencies with narrowband quantum memories, or alternatively as a quantum optical transponder, efficiently interfacing different regions of a frequency-multiplexed spectrum. 
\end{abstract}

\maketitle

   Optical frequency conversion \cite{Kumar1990} is an enabling technology for quantum computation \cite{Langford2011} and quantum communication \cite{Gisin2007}. It plays a crucial role in interfacing quantum memories, which usually work outside of the telecom spectrum (around 1550nm), with the current optical fiber infrastructure. Quantum memories \cite{Lvovsky2009} are an integral part of the quantum repeater protocol and they can be implemented in platforms such as trapped ions \cite{Wang2017}, atomic ensembles \cite{PhysRevLett.116.090501}, and rare-earth doped crystals \cite{Hedges2010}. Quantum frequency conversion has been used to demonstrate remote entanglement \cite{Duan2001} between memories separated by up to 50 km of optical fibre \cite{Yu2020}, and for the conversion of single photons emitted from trapped ions \cite{Bock2018,Walker2018} into the 1550~nm region. 
   
   Recent developments have shown potential light storage at telecom wavelengths using erbium doped materials \cite{Saglamyurek2015,Rancic2017} and engineered optomechanical systems \cite{Wallucks2020}. While compatible with the fiber network, these memories have a storage bandwidth of the order of hundreds of MHz, which is only a small fraction of the 7.2~THz available for the 72 channels in the wavelength division multiplexing (WDM) grid. This is why dynamic frequency conversion between different WDM channels plays a key role if we want exploit the full potential of the quantum fiber network via frequency multiplexing. 
   
   Conversion between WDM channels can be achieved in second and third order nonlinear materials. Third order nonlinear materials use a two pump scheme \cite{McKinstrie2005} with experiments performed in nonlinear optical fibres \cite{Clark2013,Joshi2018}, and in silicon microresonators \cite{Li2016}. The relative weakness of the $\chi^{(3)}$ nonlinear susceptibility means that these schemes require powerful pumps, very long nonlinear fibres or high quality resonators. Each of these present limits on the flexibility and bandwidth of the frequency conversion. 
   
   Second order nonlinear ($\chi^{(2)}$) materials are usually more efficient for frequency conversion, with experiments performed on periodically poled waveguides in lithium niobate \cite{Albrecht2014,Lenhard2017,Kasture2016}. Though $\chi^{(2)}$ materials are compatible with a single pump conversion scheme, this is usually limited in bandwidth because of the narrow phase-matching condition. Fortunately, it is possible to emulate the two pump $\chi^{(3)}$ scheme by cascading two $\chi^{(2)}$ frequency conversion processes. This involves performing two concurrent frequency conversions between distant wavelengths to achieve a small frequency shift. Originally suggested for tuneable conversion of classical signals \cite{Jun2003}, it was later demonstrated how a single waveguide device can perform frequency conversion between any pair of channels in the WDM grid with an average efficiency $>$80\% \cite{Fisher2020}. 
    
    Here, we demonstrate the scheme presented in \cite{Fisher2020} for high efficiency frequency shifting between WDM channels on single photons. This protocol has the dual advantages of two orders of magnitude higher conversion efficiency, and tuneability across the full telecom bandwidth when compared with similar, simplified techniques \cite{PhysRevA.97.063810}. We achieve a conversion efficiency of up to $55\pm8$\% and measured quantum interference between converted photons with a HOM dip visibility of 49.3\% (84.5\% when noise is subtracted). We also present solutions to current limitations of our set-up and for future practical applications. 

    \begin{figure*}[t]
        \includegraphics{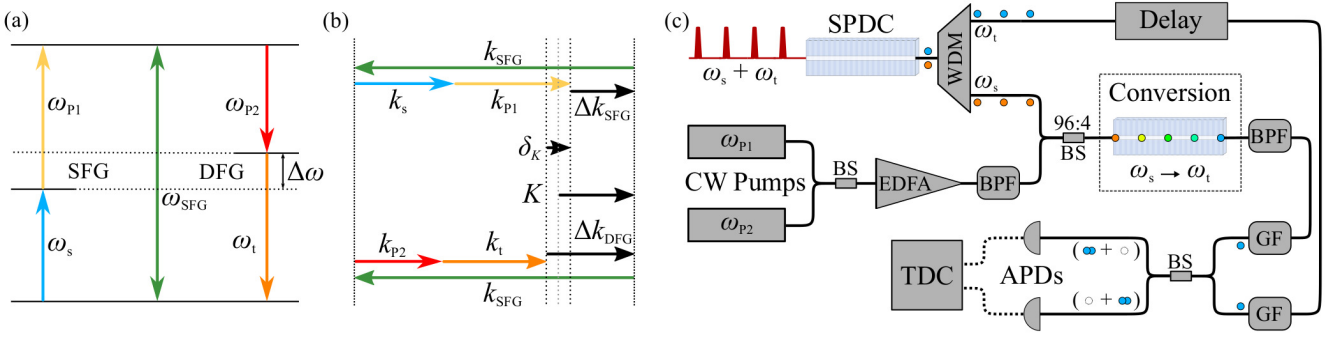} 
        \caption{\label{fig:ProtocolSetup} \textbf{(a)} Interaction between the frequency components ($\omega$). Frequency shift $\Delta\omega = \omega_\text{t}-\omega_\text{s} = \omega_\text{P1} - \omega_\text{P2}$. \textbf{(b)} The difference in wavenumber ($k$) between all the frequency components creates two phase mismatches ($\Delta k_{SFG}$ and $\Delta k_{DFG}$) for the two cascaded processes. Includes the average phase mismatch $K$ and their difference $\delta_K$. \textbf{(c)} Schematic layout of the experiment. Most parts realised using bulk optical components. Beam splitter (BS), Erbium Doped Fibre Amplifier (EDFA), Band Pass Filter (BPF), Grating Filter (GF), Avalanche Photodiodes (APDs), Time to Digital Converter (TDC), continuous wave pump lasers (CW Pumps).}
    \end{figure*}


    Our frequency conversion scheme is illustrated in Fig.~\ref{fig:ProtocolSetup}(a) and consists of concurrent stages of sum frequency generation (SFG) and difference frequency generation (DFG) in a nonlinear waveguide. In the SFG stage, the photon at the signal frequency is upconverted to the intermediate frequency by pump 1, then in the DFG stage pump 2 downconverts the photon from the intermediate frequency to the target frequency. The difference in phase velocities between the interacting modes results in the phase mismatches 
    \begin{align}
        \Delta k_{SFG}=k_{SFG}-k_{s}-k_{P1}\\
        \Delta k_{DFG}=k_{SFG}-k_{t}-k_{P2}
    \end{align}
    for the SFG and DFG processes (see Fig~\ref{fig:ProtocolSetup}(b)), where $k_x$ indicates the wavevector of a mode in the waveguide and subscript stands for s for signal, t for target, and P1 and P2 for the two pumps. These two phase mismatches can be partially compensated by quasi-phase-matching in materials that are compatible with periodic poling. It was shown in \cite{Fisher2020} that the optimal poling period for this cascaded interaction is the one that compensates for the average phase mismatch $K=(\Delta k_{DFG}+\Delta k_{SFG})/2$, while the difference $\delta_K=\Delta k_{DFG}-\Delta k_{SFG}$ determines the limit to the overall conversion efficiency.

    To describe the single photon frequency conversion, we extend the typical formulation of quantum frequency conversion \cite{Kumar1990,McKinstrie2005}, where pumps are treated as classical fields and phase mismatches are included using a rotating wave approximation, and begin with a Hamiltonian,
    \begin{align}
        H = \mathcal{X}_1E_1\hat{a}_\text{s}\hat{a}^\dagger_\text{SFG} +& \mathcal{X}_1E^*_1\hat{a}^\dagger_\text{s}\hat{a}_\text{SFG} + \mathcal{X}_2E_2\hat{a}_\text{t}\hat{a}^\dagger_\text{SFG} + \nonumber \\ \mathcal{X}_2E^*_2\hat{a}^\dagger_\text{t}\hat{a}_\text{SFG} 
        -&\frac{\delta_K}{2}\hat{a}^\dagger_\text{s}\hat{a}_\text{s} + K\hat{a}^\dagger_\text{SFG}\hat{a}_\text{SFG} + \frac{\delta_K}{2}\hat{a}^\dagger_\text{t}\hat{a}_\text{t} \;.
    \end{align}
    $\mathcal{X}_1$ and $\mathcal{X}_2$ are the coupling strengths of the SFG and DFG processes, $E_1$ and $E_2$ are the electric field strengths of pump 1 and pump 2, and $\hat{a}^\dagger_\text{s,SFG,t}$ and $\hat{a}_\text{s,SFG,t}$ are the creation and annihilation operators for the signal, sum frequency and target modes.
    Time evolution inside a waveguide is equivalent to spatial propagation, giving us a set of spatial Heisenberg equations of motion for the annihilation operators,
    \begin{subequations}
    \label{ann}
    \begin{align}
        \frac{d\hat{a}_\text{s}}{dz}=-i\frac{\delta_K}{2}\hat{a}_\text{s}+i\mathcal{X}_1E^*_1\hat{a}_\text{SFG}, \label{ann1}\\ 
        \frac{d\hat{a}_\text{SFG}}{dz}=i\mathcal{X}_1E_1\hat{a}_\text{s}+iK\hat{a}_\text{SFG}+i\mathcal{X}_2E_2\hat{a}_\text{t}, \label{ann2}\\ 
        \frac{d\hat{a}_\text{t}}{dz}=i\mathcal{X}_2E^*_2\hat{a}_\text{SFG}+i\frac{\delta_K}{2}\hat{a}_\text{t}. \label{ann3}
    \end{align}
    \end{subequations}
    Under the conditions that the poling period is given by $\Lambda=2\pi/K$ and the pump powers are tuned so that $\mathcal{X}^2_1\left|E_1\right|^2=\mathcal{X}^2_2\left|E_2\right|^2$, we can solve Eqs.~\ref{ann} and find the conversion efficiency,
    \begin{equation}
        \eta = \frac{\braket{\hat{a}^\dagger_\text{t}(L) \hat{a}_\text{t}(L)}}{\braket{\hat{a}^\dagger_\text{s}(0)\hat{a}_\text{s}(0)}} = \frac{16Q^2}{(\delta^2_K+4Q)^2}\sin^4 \left(\frac{L}{4}\sqrt{\delta^2_K+4Q}\right),
    \end{equation}
    for a waveguide of length $L$ where $Q = 2\mathcal{X}^2_1\left|E_1\right|^2 = 2\mathcal{X}^2_2\left|E_2\right|^2$.
    The maximum possible conversion efficiency obtained when $L/4\sqrt{\delta^2_K+4Q}=\pi/2+m\pi$ and is given by,
    \begin{equation}
    \label{etamax}
        \eta_\text{max} = \left(1-\frac{\delta^2_KL^2}{4\pi^2}\right)^2
    \end{equation}
    for $m=0$ where the condition $\delta^2_KL^2 \leq 4\pi^2(1+2m)^2$ has to be satisfied for all $m$.
    
\begin{figure}
        \includegraphics{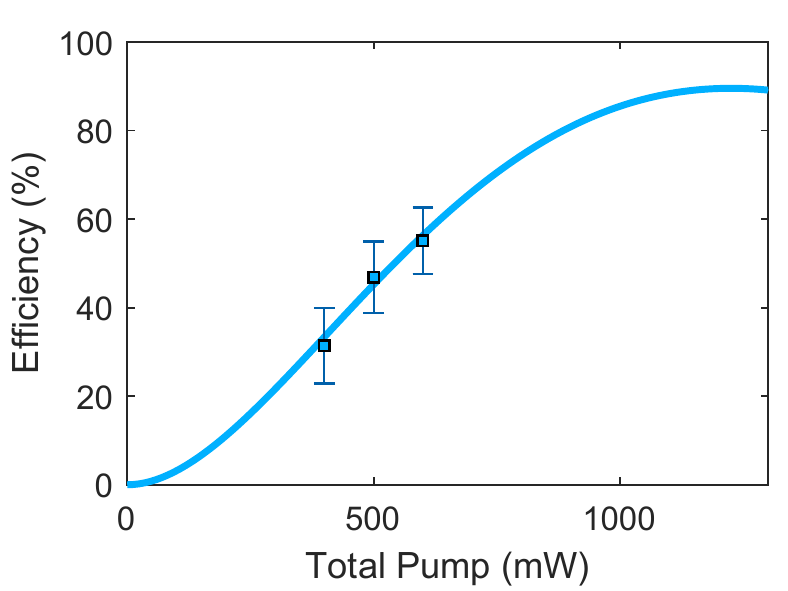} 
        \caption{\label{fig:Eff}Conversion efficiency for three different pump powers measured as the difference between counts with photons blocked and unblocked. Normalised by count rate with no conversion. Solid line is the theoretical efficiency with maximum at P=1.2~W.}
    \end{figure}

    The $\chi^{(2)}$ nonlinearity, present in non-centrosymetric materials like lithium niobate, is many orders of magnitude larger than the $\chi^{(3)}$ nonlinearity. Consequently, shorter devices using weaker pumps can achieve the same conversion efficiency as several metres of nonlinear fibre using powerful pumps. Comparing our 3.8~cm waveguide to a nonlinear fibre 100~m long \cite{Joshi2018}, we estimate maximum conversion with 1.2 W of total pump power (see Fig.~\ref{fig:Eff}) while the fibre requires 20 W. Maximum conversion can be reached with less power in a much longer nonlinear fibre (300~mW in 750~m \cite{Clark2013}), but as Eq.~\ref{etamax} shows, extra length is detrimental to maximum conversion efficiency. This is also true since the phase mismatch difference $\delta_K$ is the same as the phase mismatch for the $\chi^{(3)}$ process, with similar effect to maximum efficiency shown in Eq.~\ref{etamax}.

    A schematic of the experimental setup is shown in Fig.~\ref{fig:ProtocolSetup}(c). The photon pairs were generated by SPDC in a reverse-proton-exchanged periodically poled lithium niobate (RPE:PPLN) waveguide \cite{Lenzini2015} using a erbium femtosecond fiber laser with a repetition rate of 25~MHz. A 100~GHz portion of the laser spectrum, centred at 1537.40~nm (corresponding to channel 50 of the dense WDM grid), was selected by a WDM module, amplified, frequency doubled, and used to generate spectrally separated photon pairs around 1537.40~nm. The average second harmonic power was 20~{\textmu}W, corresponding to a pair generation probability per pulse of $\sim$1-2\% and a coincidence to accidental ratio of 8.55. The photons were separated in a WDM module, with signal photons at 1538.98~nm (channel 48) sent to be converted and idler photons at 1535.82~nm (channel 52) sent to a delay stage. Both before and after the SPDC more than 150~dB of filtering was used to ensure clean photon pair generation.  
    
    The pumps used for the frequency conversion of the signal photons were two tunable CW diode lasers with pump 1 set to 1550.28~nm and pump 2 set to 1553.49~nm, 400~GHz lower in frequency to make the +400~GHz shift in the converted light to match the frequency of the target single photon. Both lasers were passed thru a single erbium-doped fibre amplifier and their output powers balanced to within 0.2\% for optimal conversion.  Following amplification to a combined power of 16 W, these pumps passed through approximately 90~dB of bandpass filtering to eliminate any amplifier noise near the photon frequencies. Using polarization optics, the pumps were then combined with the signal photons from SPDC at a ratio of about 96:4, favouring the single photons. This left approximately 600~mW of pump power for conversion and allowed the pump power to be tuned using a half-wave plate, with only a small penalty to photon losses.
    
    The combined pumps and single photons were then coupled to another RPE:PPLN waveguide for frequency conversion. The frequency conversion device was 3.8 cm long and had a normalized conversion efficiency for second harmonic generation of 46~\%W$^{-1}$cm$^{-2}$. It was kept at a temperature of 122.5~{\degree}C for optimal phase matching and to avoid high-power photorefractive effects. The facets of the chip were anti-reflection coated with a reflectivity less than 1\%.
    
    After conversion the single photons were separated from the pumps by a band-pass filter (Semrock 1550/3) with the reflected pumps used to monitor the coupled power. A diffraction grating (Thorlabs 1200/mm holographic) was used to further suppress the amount of pump light present in the converted photon spatial mode. The grating was operated close to the Littrow angle, sending converted photons back through the bandpass filter to a mirror, then through the filter again. The transmisson spectrum of this filtering stage was 28.6~GHz FWHM centred at 1535.82~nm. A similar filter set-up was incorporated into the delay stage to maximise the spectral overlap between converted and unconverted photons.
    
    The single photons were detected using two InGaAs single photon detectors (SPD, IDQuantique ID200 and ID201) gated with a 2.5~ns window at 1~MHz rate and synchronized with the laser pulses. Coincidences were counted using a time to digital converter (UQDevices Logic16). Figure~\ref{fig:Eff} shows the conversion efficiency as a function of the total pump power which reaches  $55\pm8$\% for a total pump power of 600~mW. The efficiency was measured as the ratio between single counts of converted photons by the SPD, versus single counts of unconverted photons with the pumps off and the filters tuned to 1538.98~nm.  Noise counts from the pumps were subtracted by blocking the single photons before the waveguide. The theoretical fit in Fig.~\ref{fig:Eff} shows that our waveguide requires a total pump power of approximately 1.2 W for a theoretical maximum conversion of 89.6\%. The fit includes the effects of waveguide propagation loss and the wrong pump interaction. 
      \begin{figure}
        \includegraphics{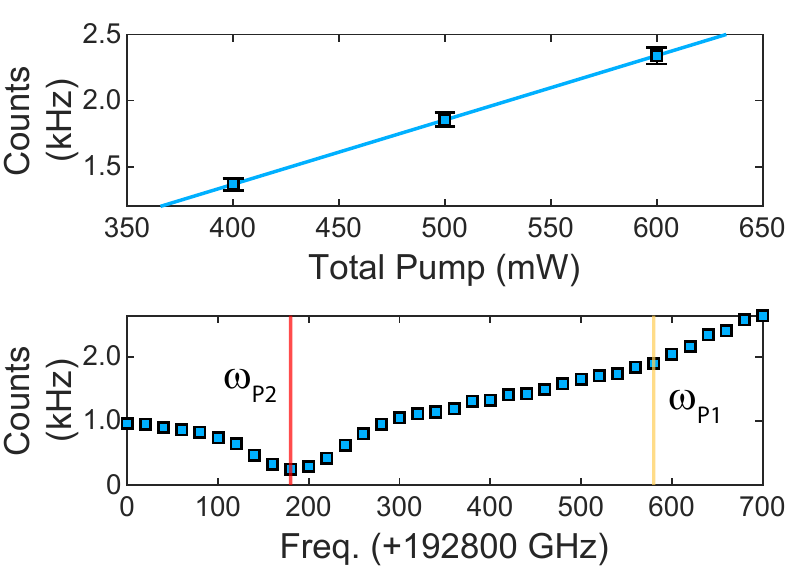}
        \caption{\label{fig:Noise}\textbf{(a)} Noise counts for the total pump power $P_1+P_2$. \textbf{(b)} Noise spectrum from the conversion waveguide with a single pump at 500~mW. Vertical bars indicate chosen pump frequencies which minimised combined noise.}
    \end{figure}
    
    The high efficiency $\chi^{(2)}$ process allows undesirable interactions between pumps and signal, primarily introducing an effective loss through signal photons interacting with the wrong pump. The pumps themselves may also generate second harmonics or undergo SFG between themselves, and these new frequencies can create noise photons by SPDC. Although these interactions are always present, their effects can be reduced by choosing pumps further away from the signal and target frequency as described in \cite{Fisher2020}.
    
    The primary source of noise is Raman scattering. This is supported by the linear increase in noise with input power shown in Fig.~\ref{fig:Noise}a. Since the total amount of combined second harmonic and sum frequency generated by 500 mW of pump is approximately 2~mW, these process and any subsequent SPDC are poorly phase matched and very inefficient. Figure~\ref{fig:Noise}b shows the noise spectrum near our possible choices of pumps frequencies. Considering that the pumps have to be 400~GHz apart, only marginal improvement can be achieved tuning their frequencies. Finite temperature tuning range of our phase matching limited operation to this region. 
    
\begin{figure}
        \includegraphics{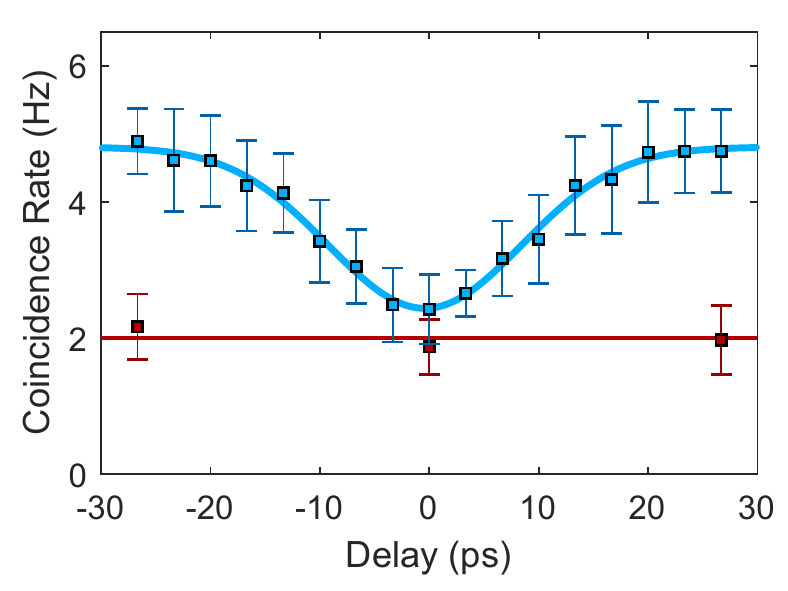}
        \caption{\label{fig:HOM}Blue curve is the Hong-Ou-Mandel interference dip between converted and unconverted photons. Red line is the constant noise floor measured as the combined rate of accidental coincidences between noise only, unconverted only and noise, and converted only and noise.}
    \end{figure}

    Indistinguishability between the converted and the unconverted photon was verified with a Hong-Ou-Mandel (HOM) interference \cite{PhysRevLett.59.2044} measurement in a fibre beam splitter, converting the photons with a total pump power of 500~mW. A tuneable delay was added to the unconverted (target) channel and the number of coincidences after the beamsplitter was measured as a function of delay . Figure~\ref{fig:HOM} shows the resulting HOM dip with a raw visibility of $49.3\pm2.7$\%, limited by the noise from uncorrelated Raman photons. Noise subtracted visibility is $84.5\pm4.6$\% which is limited by the imperfect 49.3:50.7 splitting ratio of the fiber splitter and the spectral overlap between the two filters of 96\%.
    
     Uncorrelated Raman photons are the main limiting factor of our device. Several approaches can be used to mitigate this problem including using pump wavelengths around 2.4~{\textmu}m, which also allow for efficient conversion \cite{Fisher2020}. This is because the telecom region lies on the opposite side of an inflection in the refractive index curve of lithium niobate to 2.4~{\textmu}m. Consequentially, this allows the difference in phase mismatches $\delta_K$ to be even smaller than what is allowed with telecom pumps, improving efficiency. Moving to picosecond pulsed pumps with better time gating of the detectors and heralded generation of the single photons can also be implemented to reduce the noise by at least two orders of magnitude.
    
    In conclusion, we have demonstrated a scheme for the conversion of single photons between telecom WDM channels using cascaded sum and difference frequency generation. Using our waveguide, we have achieved a conversion efficiency up to $55\pm8$\% and shown quantum interference between converted photons. Although Raman generation in the waveguide reduced the visibility of the HOM dip, experimental improvements around the pump lasers can be used to mitigate this effect. This scheme will allow frequency mulitplexed quantum communication across the entire telecom spectrum to be efficiently interfaced with narrowband rare-earth quantum memories, potentially enabling long distance high bandwidth quantum communication.
    \\
    
    This work was supported by the Australian Research Council (ARC) Centre of Excellence for Quantum Computation and Communication Technology (CE170100012), and the Griffith University Research Infrastructure Program. ML was supported by the Australian Research Council (ARC) Future Fellowship (FT180100055). PF was supported by the Australian Government Research Training Program Scholarship. BH was supported by the Griffith University Postdoctoral Fellowship. This work was performed in part at the Queensland node of the Australian National Fabrication Facility, a company established under the National Collaborative Research Infrastructure Strategy to provide nano- and microfabrication facilities for Australia's researchers.
    
\bibliography{qfcbib}

\end{document}